\documentclass[letterpaper, 10 pt, conference]{ieeeconf}  

\IEEEoverridecommandlockouts                              
\usepackage{amsmath} 
\usepackage{amssymb}  
\usepackage{subcaption}
\usepackage{amsfonts}
\usepackage{amsthm}
\usepackage{comment}
\usepackage{tikz}
\usetikzlibrary{graphs,quotes,arrows,positioning}
\newcommand{\E}{\mathbb{E}}
\newcommand{\Var}{\mathbf{Var}}

\newtheorem{theorem}{Theorem}
\newtheorem{lemma}[theorem]{Lemma}
\newtheorem{fact}[theorem]{Fact}

\title{\LARGE Rule Designs for Optimal Online Game Matchmaking
}

\author{Mingkuan Xu, Yang Yu, \emph{and} Chenye Wu
	\thanks{The authors are with the Institute for Interdisciplinary Information Sciences, Tsinghua University, Beijing, 100084, P.R. China.  C. Wu is the correspondence author. Email: chenyewu@tsinghua.edu.cn.}
	\thanks{This work was supported in part by Turing AI Institute of Nanjing.}
}

\begin{document}

	\maketitle
	\thispagestyle{empty}
	\pagestyle{empty}

	\begin{abstract}
		
		Online games are the most popular form of entertainment among youngsters as well as elders. Recognized as e-Sports, they may become an official part of the Olympic Games by 2020. However, a long waiting time for matchmaking will largely affect players' experiences. We examine different matchmaking mechanisms for 2v2 games. By casting the mechanisms into a queueing theoretic framework, we decompose the rule design process into a sequence of decision making problems, and derive the optimal mechanism with minimum \emph{expected} waiting time. We further the result by exploring additional \emph{static} as well as \emph{dynamic} rule designs' impacts. In the static setting, we consider the game allows players to choose sides before the battle. In the dynamic setting, we consider the game offers multiple zones for players of different skill levels. In both settings, we examine the value of choice-free players. Closed form expressions for the expected waiting time in different settings illuminate the guidelines for online game rule designs.
		
	\end{abstract}
	
	\vspace{0.1cm}
	
	\section{Introduction}
	
	Long matchmaking waiting time can be a disaster for online games. However, such situation is very common to newborn or obsolete games with few players. Even players with high ranks in popular games may experience long waiting times, sometimes longer than the game time. For example, public data from a \emph{League of Legends} server \cite{ref_lol} shows that the average waiting time reaches up to 45 minutes for the highest ranks, while \emph{League of Legends} is a session-based game with relatively short sessions, averaging around 34 minutes according to the game developers \cite{ref_trace}.
	
	Conventional wisdom may suggest attracting more (high quality) players to the game, which often involves huge investments on game designs. While such investments certainly help, we focus on the rule design of games, which is the other determinants of the game's attractiveness. 
	We submit that a good rule design can significantly improve the game's attractiveness without any huge extra investment.
	
\begin{figure}[t]
		\centering
		\includegraphics[width=2.8in]{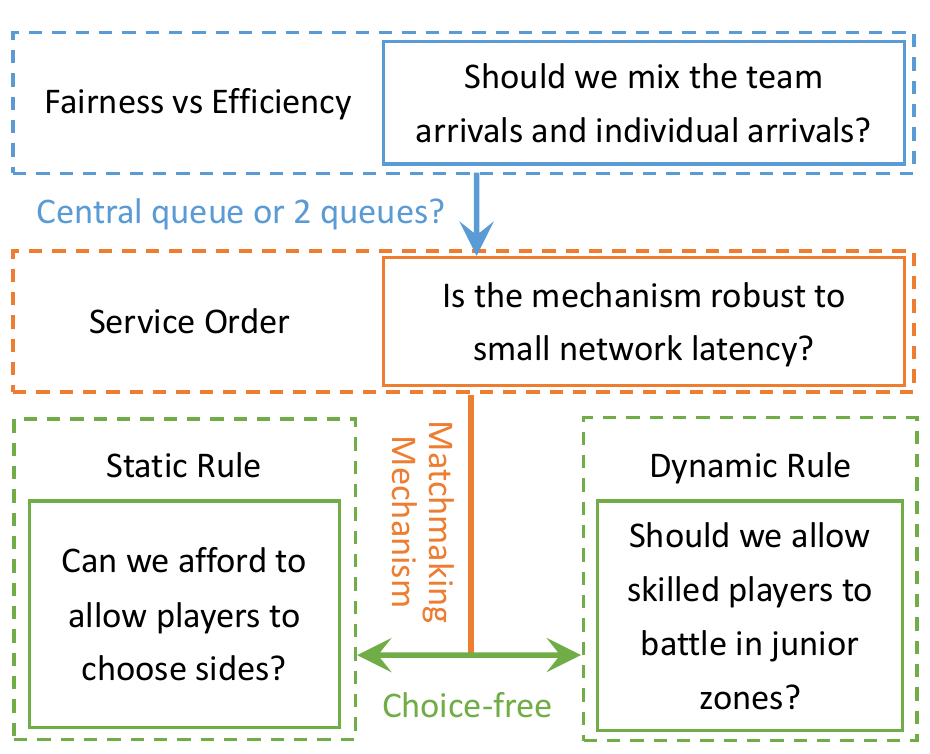}
		\caption{The rule design for a 2v2 battle game.\vspace{-0.5cm}}
		\label{fig:rule_design}
	\end{figure}

	Figure \ref{fig:rule_design} illustrates our decomposition of the whole rule chain into a sequence of \emph{decision making} problems. We sequentially discuss the rule design of each component in the rule chain and their combinations. Using a queueing theoretic framework, we seek to answer these \emph{decision making} problems and offer more insights on better rule designs for online games. 
	
	\subsection{Related Work}
	
	Most previous research on online game matchmaking focuses on understanding the network latency's impact. For example, Agarwal \emph{et al.} conduct the latency estimation for matchmaking through geolocation and network coordinate systems in \cite{ref_latency_sensitive}. Manweiler \emph{et al.} extend the research to mobile games in \cite{ref_switchboard}, where the network is much more unstable.
	
	Surprisingly, the mathematical treatment for online game matchmaking waiting time only emerges recently. Vron \emph{et al.} study user traces and analyze the influence of ranking on the waiting time in \cite{ref_trace}. Yuval \emph{et al.} propose to define the matchmaking cost regarding the matched requests and the waiting time, and study the minimal cost perfect matching in \cite{ref_haste}.  Malak and Deja examine the game structure's impact on matchmaking in \cite{ref_game_structure_sensitive}. Alman and McKay lay out theoretical foundations for matchmaking, and propose a solution to extract the fairest and most uniform games in \cite{ref_foundation}.
	
	
	\subsection{Our Contributions}
	 In contrast to previous works, we focus on understanding the waiting time through a queueing framework, and we seek to understand different rule designs' impact on the matchmaking mechanism. The principal contributions in this paper can be summarized as follows:
	 
	 \begin{itemize} \setlength\itemsep{0.3em}
	 \item \emph{Decomposition of Rule Design}: Based on the queueing framework, we decompose the whole rule chain design into a sequence of decision making problems. Such decomposition allows us to discuss the rule design of each component and their combinations, and thus to identify bottleneck in rule designs. 
	     \item \emph{Optimal Matchmaking Mechanism}: We use 2v2 battle game to derive the optimal matchmaking mechanism. By comparing different service orders for diverse arrivals, we submit that the optimal matchmaking mechanism is the \emph{packing} service order and it is robust to small network latency.
	     \item \emph{Static Rule Design}: We use choosing sides before the battle as an example to highlight the impact on static rule design to the matchmaking mechanism. We find that allowing all players to choose sides yields \emph{unbounded} expected waiting time. This inspires us to examine the value of choice-free players for the matchmaking mechanism. We show that a small portion of choice-free players can already guarantee a reasonable expected waiting time.
	     \item \emph{Dynamic Rule Design}: As for the dynamic rule design, we consider the game offers multiple zones for players of different levels, and examine whether the game designer should allow skilled players to battle in a junior zone. Coincidentally, we refer to the skilled players who are indifferent in the battle zones as choice-free players. Being choice-free benefits themselves as well as the whole system.
	 \end{itemize}
	 
	 \vspace{0.1cm}
	The rest of the paper is organized as follows. Section \ref{sec:systemModel} introduces our system model and revisits the basic queueing results for 2v2 battle games. By comparing different service orders, we derive the optimal matchmaking mechanism in Section \ref{sec:design}. Based on this optimal mechanism, we investigate the impacts of \emph{static} and \emph{dynamic} rule designs in Section \ref{sec:static} and Section \ref{sec:dynamic}, respectively. Section \ref{sec:conclusion} delivers the concluding remarks and points out possible future directions.

	\section{System Model}
	\label{sec:systemModel}
	
	The matchmaking problem can be modeled in the queueing framework. When a player starts to search for a game, it joins the queueing system. Take 2v2 game as an example: players can make a team with a friend and join the queueing system together, or choose to join the system on their own. Through the matchmaking mechanism, when there are sufficient players (in the 2v2 game setting, four players are sufficient), they may start a game and leave the queue. 
	
	\subsection{Assumptions}
	To simplify our analysis and establish a stylized model, we make the following assumptions:
	
	\begin{enumerate}
	    \item The network latency is negligible.
	    \item The arrival process of players follows Poisson Process.
	    \item All players are identical regarding their skills.
	\end{enumerate}
	
	\vspace{0.3cm}
	\noindent \textbf{Remark}: We will relax the first assumption by examining the matchmaking mechanism's robustness to different service orders. The last assumption guarantees that the players are indifferent to their teammates as well as their rivals. In many games, only players with the similar ranks can be matched together, and our subsequent analysis can be generalized to such cases straightforwardly. For other games, they may need to take into account players' self achievement when designing the matchmaking, which is outside the scope of this paper. This assumption allows us to focus on understanding the waiting time in different matchmaking mechanisms, and our conclusions can serve as the benchmark for general online game matchmaking mechanism design.
	
	\vspace{0.2cm}
	In order to facilitate our understanding of the rule designs' impact on matchmaking mechanisms, in this section, we review the classical queueing theory results for two cases: the $k$-player game, and the 2v2 battle game. 
	
	\subsection{$k$-player Games}
	
	The first type of game is the most basic one, standard $k$-player game. One example could be Mahjong game, which involves four identical players. The simple matchmaking mechanism waits for a total of $k$ players and starts a game.
	
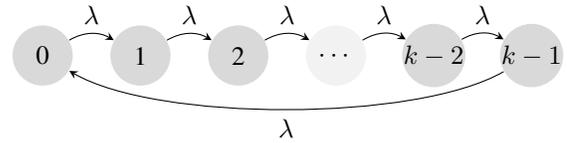
\begin{figure}[t]
		\centering
		\begin{tikzpicture}
		\graph[->/.tip=stealth,
		grow right=1.3cm,
		nodes={circle,fill=black!15,inner sep=0,minimum size=8mm}]
		{
			{
				[edges={bend left,"$\lambda$" above}]
				0 -> 1 -> 2 -> 3[as=$\cdots$,fill=black!5] -> 4[as=$k - 2$] -> 5[as = $k - 1$];
			};
			5 ->[bend left=30,looseness=0.6,"$\lambda$" below] 0;
		};
		\end{tikzpicture}
		\caption{The CTMC of the queueing system for $k$-player game. \vspace{-0.5cm}}
		\label{kplayerqueue}
	\end{figure}

	Denote the Poisson arrival rate of players by $\lambda$. Figure \ref{kplayerqueue} shows the Continuous Time Markov Chain (CTMC) of such a $k$-player queue. By solving the balance equations, we can prove the following lemma.
	\begin{lemma}
	    For $k$-player games with Poisson arrival rate $\lambda$, any matchmaking mechanism, which ensures that players start a game as soon as there are at least $k$ of them in the queue, yields the same
	    expected waiting time $\E[T]$ as follows:
    	\begin{equation}
    		\E[T]^{\text{$k$-player}} = \frac{k - 1}{2 \lambda}.
    	\end{equation}
    	\label{thm_k_player}
	\end{lemma}
    \noindent \textbf{Proof}: 	It is sufficient to identify that the stationary probability distributions are as follows
	\begin{equation}
	    \pi_0 = \pi_1 = \cdots = \pi_{k - 1} = \frac{1}{k}.
	\end{equation}
	Then, Little's Law \cite{ref_little} immediately yields the result. \hfill $\blacksquare$
	
	\vspace{0.2cm}
	\noindent \textbf{Remark}: When the game only requires a single player, i.e., $k=1$, our conclusion also holds. It simply degenerates to the trivial case where there won't be any queue in the system.
	
	
	\subsection{2v2 Games}
	Another popular form of game involves battles, and can often be modeled as a 2v2 game. One popular example could be \textit{Clash Royale} \cite{ref_cr}. In such battle games, players can make a team with a friend to join the battle, or just go on a quick match by themselves.
	
	Denote the individual arrival rate by $\lambda_1$, and the team (consisting of 2 players) arrival rate by $\lambda_2$. Thus, the total arrival rate of players into the queueing system is $\lambda_{\text{total}} = \lambda_1 + 2\lambda_2$.
	
\begin{figure}[t]
		\centering
		\begin{tikzpicture}
		\graph[->/.tip=stealth,
		grow right=2cm,
		nodes={circle,fill=black!15,inner sep=0,minimum size=8mm}]
		{
			{
				[edges="$\lambda_1$"]
				{
					[edges={above, bend left}]
					0 -> 1 -> 2 -> 3;
				};
				{
					[edges={below, bend left=50, looseness=0.9}]
					3 -> 0;
				};
			};
			{
				[edges="$\lambda_2$"]
				{
					[edges={above, bend left=60, looseness=0.9}]
					0 -> 2;
					1 -> 3;
				};
				{
					[edges={below, bend left=30}]
					2 -> 0;
					3 -> 1;
				};
			};
		};
		\end{tikzpicture}
		\caption{The CTMC of a 2v2 queue.\vspace{-0.5cm}}
		\label{2v2queue}
	\end{figure}
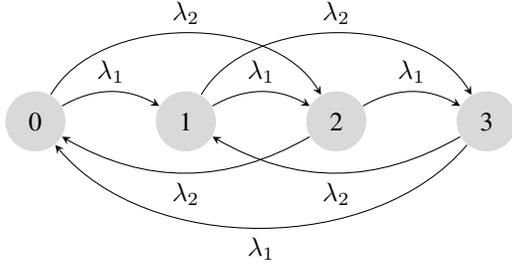

	We seek to understand the average waiting time, denoted by $T$, for all kinds of players. By analyzing the CTMC as shown in Fig. \ref{2v2queue}, we can prove the following Lemma.
	
	\begin{lemma}
	    For 2v2 games with total arrival rate $\lambda_{\text{total}}$ with $\lambda_1>0$, any matchmaking mechanism, which starts a game as soon as there are at least four players in the queue, enjoys the same expected waiting time $\E[T]$ as follows
    	\begin{equation}
    		\E[T]^{\text{2v2}} = \frac{3}{2 \lambda_{\text{total}}}.
    	\end{equation}
    	\label{thm_2v2}
	\end{lemma}
	\noindent \textbf{Remark}: When $\lambda_1 = 0$, the 2v2 battle reduces to the 2-player case. In this case, Lemma \ref{thm_2v2} won't hold. Hence, we require the additional condition $\lambda_1>0$ to maintain the structure of the 4-player game without degeneration.
	
	\subsection{Criteria for Rule Design}
	The rule design must consider the associated system performance measured by the total expected waiting time, and individual impacts assessed by the variance of waiting time.
	While Lemma \ref{thm_2v2} is remarkable in guaranteeing the same expected waiting time for all players, different matchmaking mechanisms may yield diverse expected waiting time for individual arrivals and the team arrivals. This leads us to employ the variance of the waiting time for all arrivals as the evaluation metric for different matchmaking mechanisms before the static or dynamic designs are involved. Once the rule designers begin to trade-off their choices for the static and dynamic rule designs, the conditions of Lemma \ref{thm_2v2} no longer hold. Thus, both the system performance and individual impacts are included as criteria for rule selection. 
	
	\section{Rule Design: the Basics}
	\label{sec:design}
	
	
	Experts in queueing theory may propose the most straightforward mechanism is First-In-First-Out (FIFO) central queue system. However, from player's point of view, we submit the fairest but seemingly inefficient mechanism is to divide individual arrivals and team arrivals into 2 different queues. In this section, we first compare the central queue with separate queue systems, then evaluate the performance of matchmaking mechanisms with various service orders.
	
	\subsection{Central Queue or Two Queues?}
	
	
	 The two-queue system is fair to players in that a team of friends usually cooperate better than a team of 2 random players. And the team players may even collude offline.
	 
	
	We denote the waiting time for individual arrivals by $T_1$, and that for team arrivals by $T_2$. Such a two-queue system divides the 2v2 game into a 4-player game with arrival rate $\lambda_1$ and a 2-player game with arrival rate $\lambda_2$. Lemma \ref{thm_k_player} dictates
	\begin{align}
		\E[T_1]^{\text{2Q}} &= \frac{3}{2\lambda_1}, \\
		\E[T_2]^{\text{2Q}} &= \frac{1}{2\lambda_2}.
	\end{align}
	Based on $\E[T_1]^{\text{2Q}}$ and $\E[T_2]^{\text{2Q}}$, standard mathematical manipulations yield
	\begin{align}
		\E[T]^{\text{2Q}} &= \frac{5}{2\lambda_{\text{total}}}, \\
		\Var[T]^{\text{2Q}} &= \frac{3\lambda_1^2 - \lambda_1\lambda_2 + 11\lambda_2^2}{2\lambda_{\text{total}}^2\lambda_1\lambda_2}.
	\end{align}
	
	\vspace{0.1cm}
	\noindent \textbf{Remark}: Obviously, in this straightforward yet fair mechanism, fairness pays! The expect waiting time for all kinds of arrivals is 67\% more than the class of mechanisms considered in Lemma \ref{thm_2v2}. Later, we will use numerical studies to highlight that the variance $\Var[T]^{\text{2Q}}$ is also significantly larger than many other mechanisms.
	
	\subsection{Service Order for Central Queue}
	When there is only one central queue for the 2v2 game, players are matched as soon as there are 4 players in the queue. However, even with the Poisson arrival assumption, there could be a non-negligible probability for 5 players waiting in the queue: a team arrives when 3 players are waiting in the queue. For the 3 players in queue, if there is already a team, then we need to match the two teams together for the new game since teams cannot be broken up. Hence, the remaining hurdle is to choose a service order for the case of 3 individual players in queue to conduct the matchmaking. Possible service orders include:
	\begin{enumerate} \setlength\itemsep{0.3em}
	    \item \emph{FIFO}: According to players' arrival time, match the first two individual players with the team.
	    \item \emph{Packing}: Whenever there are two individual players in the queueing system, pack them into a team. Thus, we only need to match the teams in FIFO service order. In our 2v2 game setting, it is equivalent to FIFO.
	    \item \emph{Last-In-First-Out, LIFO}: In contrast to FIFO, LIFO matches the \emph{last} two individual players with the team according to their arrival time.
	\end{enumerate}
	
	\vspace{0.1cm}
	\noindent \textbf{Remark}: It is straightforward to see that FIFO outperforms LIFO. However, it is important to consider both service orders due to \emph{network latency}. Small network latency may affect the true arrival orders of the players. Hence, evaluating both service orders' performance help us evaluate 2v2 battle game's sensitivity to service order and thus its sensitivity to small network latency.
	
	Note that Lemma \ref{thm_2v2} dictates the mean waiting time is the same for all the three service orders:
	\begin{equation}
		\E[T]^{\text{FIFO}} =\E[T]^{\text{Packing}}=\E[T]^{\text{LIFO}}= \frac{3}{2\lambda_{\text{total}}}.
	\end{equation}
	
\begin{figure}[t]
		\centering
		\begin{tikzpicture}[->/.tip=stealth,nodes={inner sep=0,minimum size=8mm}]
		\node[circle,fill=black!15] (0) at (0, 0) {0};
		\node[circle,fill=black!15] (1) at (3, 0) {1};
		\node[circle,fill=black!15] (2a) at (0, 3) {$2a$};
		\node[circle,fill=black!15] (3a) at (3, 3) {$3a$};
		\node[circle,fill=black!15] (2b) at (0, -3) {$2b$};
		\node[circle,fill=black!15] (3b) at (3, -3) {$3b$};
		\draw (0) edge[->,"$\lambda_1$" above] (1);
		\draw (1) edge[->,"$\lambda_1$" above right=-0.3, near end] (2a);
		\draw (2a) edge[->,"$\lambda_1$" above] (3a);
		\draw (3a) edge[->,"$\lambda_1$" above left=-0.3, near end] (0);
		\draw (2b) edge[->,"$\lambda_1$" above] (3b);
		\draw (3b) edge[->,"$\lambda_1$" above] (0);
		\draw (0) edge[->,"$\lambda_2$" left, bend right] (2b);
		\draw (1) edge[->,"$\lambda_2$" right, bend left] (3b);
		\draw (2a) edge[->,"$\lambda_2$" left] (0);
		\draw (3a) edge[->,"$\lambda_2$" right] (1);
		\draw (2b) edge[->,"$\lambda_2$" left, bend right] (0);
		\draw (3b) edge[->,"$\lambda_2$" right, bend left] (1);
		\end{tikzpicture}
		\caption{The CTMC for single queue 2v2 game.\vspace{-0.0cm}}
		\label{2v2FIFO}
	\end{figure}
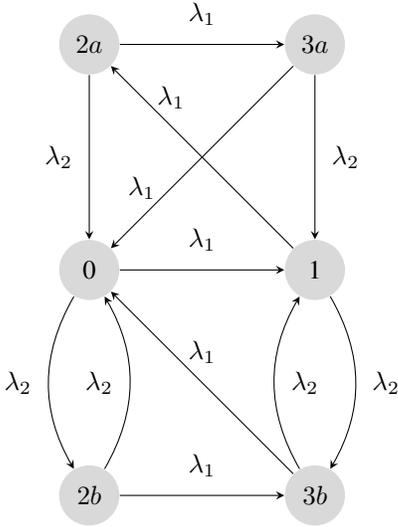

	However, the CTMC in Fig. \ref{2v2queue} does not contain enough information for us to evaluate the waiting time variance resulting from different service orders. Let states $2a$ and $3a$ denote that players in the queue are all individuals, and states $2b$ and $3b$ denote that there is a team in the queue. This leads to a CTMC in Fig. \ref{2v2FIFO}. It is worth noting that all the three service orders yield the same CTMC.
	
	By solving stationary equations for this CTMC, we can obtain desired evaluation metrics for the three service orders:
	\begin{equation}
	    \begin{aligned}
    	 &\Var[T]^{\text{FIFO}} =\Var[T]^{\text{Packing}}= \frac{7 \lambda_1^2+\lambda_1 \lambda_2+4 \lambda_2^2}{2 \lambda_1 (\lambda_1+\lambda_2)^3} \\
    	&\ \ \ \ \ \ \ \ \ \ \ \ \ \ -\frac{2 \lambda_1 \lambda_2^3+3 \lambda_2^4}{\lambda_1 \lambda_{\text{total}} (\lambda_1+\lambda_2)^4}- \left [ \E[T]^{\text{FIFO}} \right ] ^2,
    	\end{aligned}
	\end{equation}

	\begin{equation}
		\begin{aligned}
			& \Var[T]^{\text{LIFO}} = \frac{5 \lambda_1^3 \lambda_2+7 \lambda_1^2 \lambda_2^2+10 \lambda_1 \lambda_2^3+2 \lambda_2^4}{2 \lambda_1 (\lambda_1+\lambda_2)^3 \left(\lambda_1^2+2 \lambda_1 \lambda_2+2 \lambda_2^2\right)} \\
			& \ \ \ \ \ \ \ \ \ \ \ \ \ \ +\frac{8 \lambda_1^2+7 \lambda_1 \lambda_2}{2 (\lambda_1+\lambda_2)^4} - \left [ \E[T]^{\text{LIFO}} \right ] ^2.
		\end{aligned}
	\end{equation}
	
	\begin{figure}[t]
	    \centering
	    \includegraphics[width=2.6in]{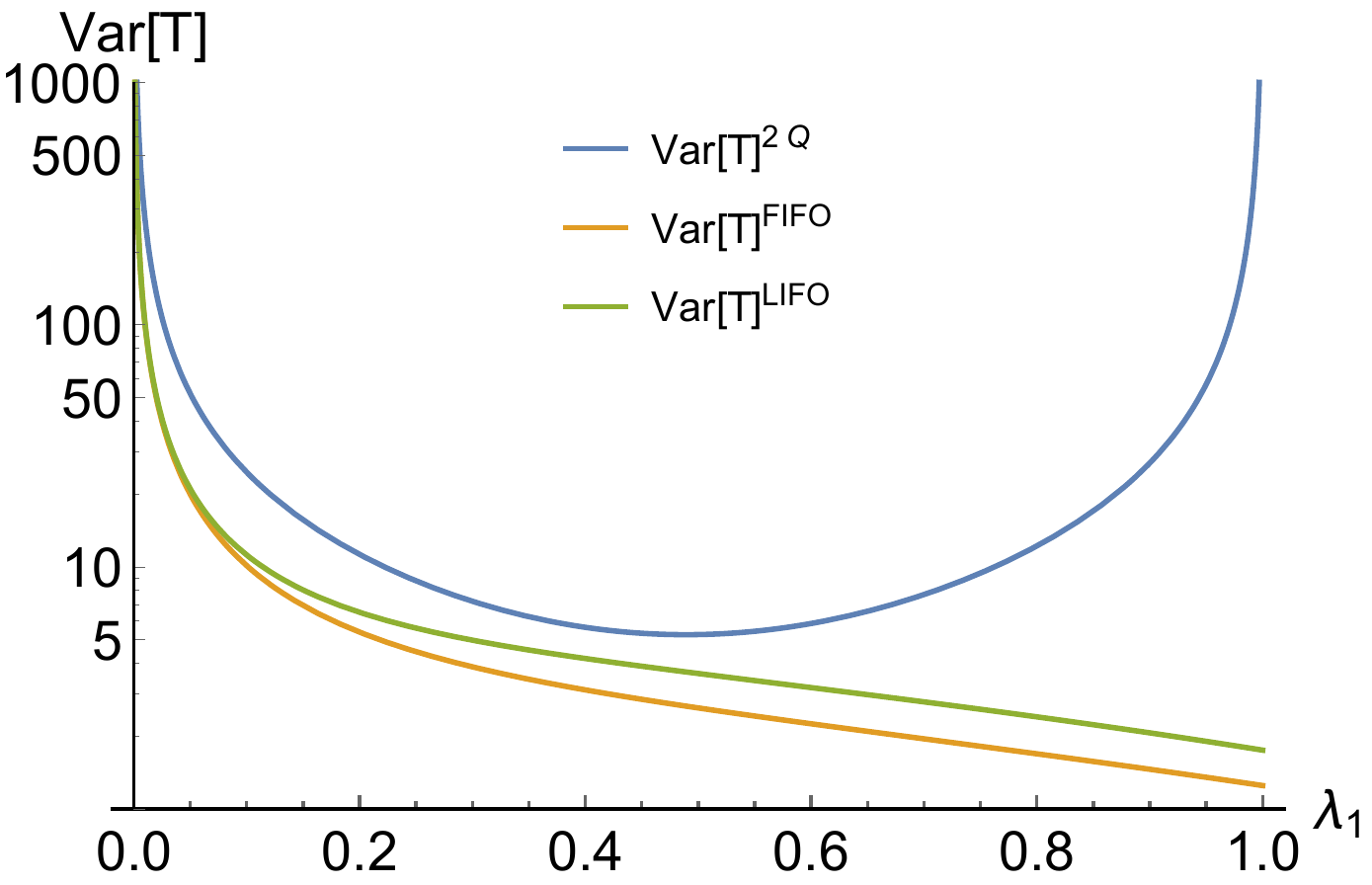}
	    \caption{The variances of waiting time with normalized $\lambda_{\text{total}}$.\vspace{-0.5cm}}
	    \label{2v2graph}
	\end{figure}
	
	
	\begin{figure*}[t]
	    \centering
	    \begin{subfigure}{.3\textwidth}
  \centering
  \includegraphics[width=1.\linewidth]{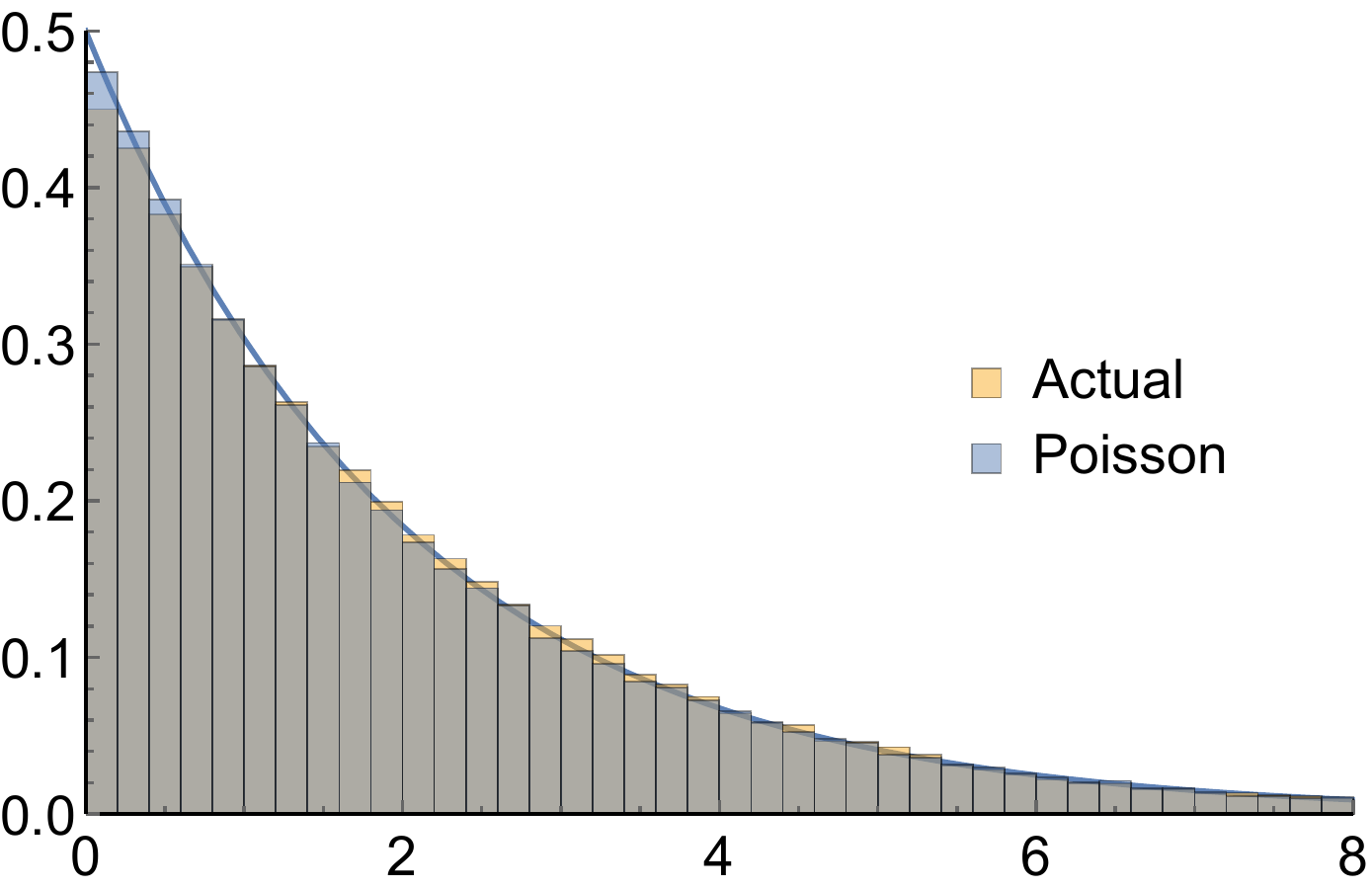} 
  \caption{$\lambda_1 = 0.2$}
  \label{fig:sub-first}
\end{subfigure}
\begin{subfigure}{.3\textwidth}
  \centering
  \includegraphics[width=1.\linewidth]{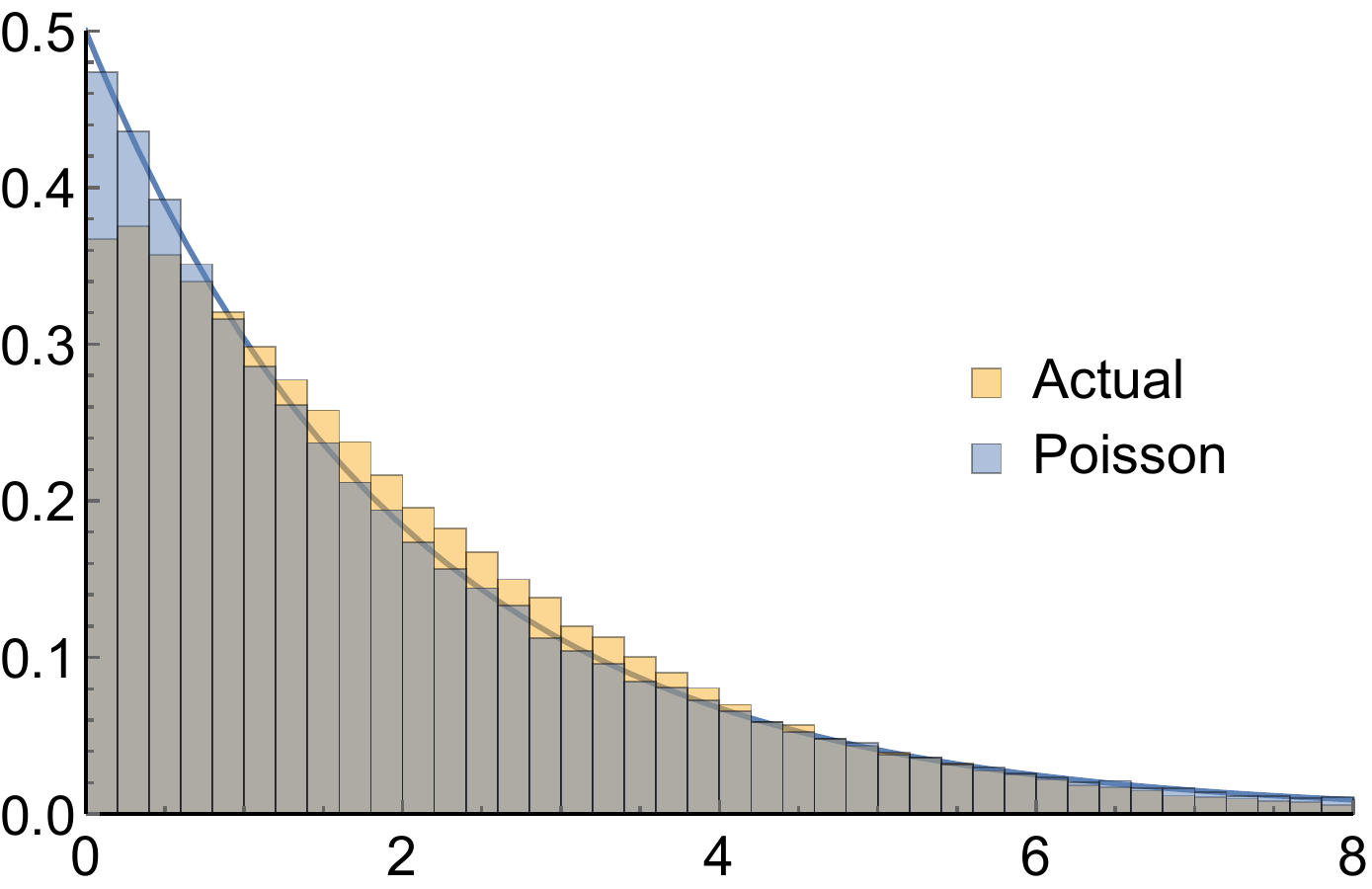}  
  \caption{$\lambda_1= 0.5$}
  \label{fig:sub-second}
\end{subfigure}
\begin{subfigure}{.3\textwidth}
  \centering
  \includegraphics[width=1.\linewidth]{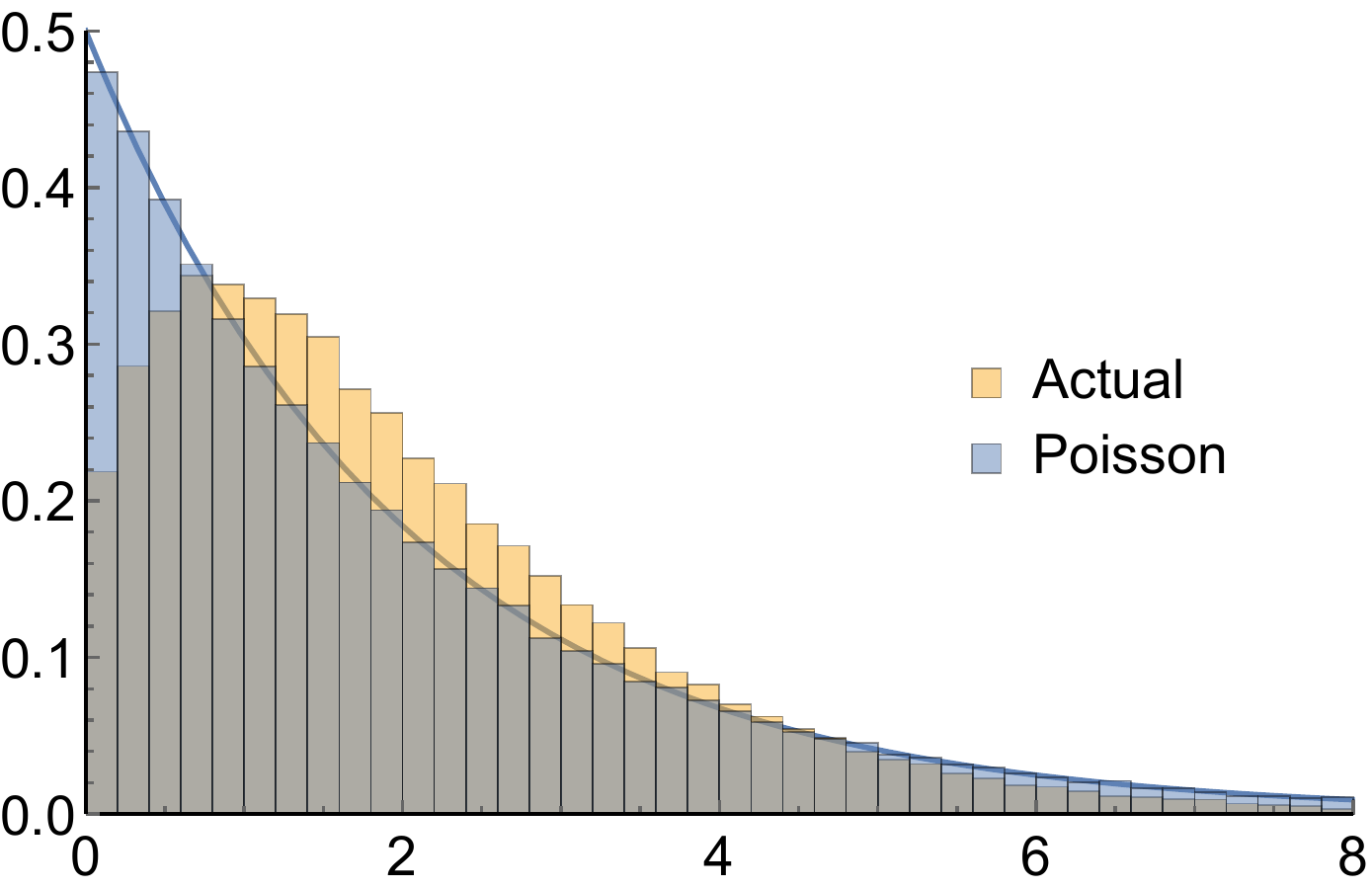}  
  \caption{$\lambda_1 = 0.8$}
  \label{fig:sub-third}
\end{subfigure}
\caption{Evaluate the inter-arrival time of 1v1 game approximation with diverse $\lambda_1$'s with normalized $\lambda_{\text{total}}$ (orange bins: histograms for 1v1 game approximation, blue bins: histograms for Poisson process with the same total arrival rate). }
\label{fig:approx}
	\end{figure*}
	
	\subsection{Performance Assessment and Rule Design}
	 Figure \ref{2v2graph} compares the variance of different matchmaking mechanisms by normalizing $\lambda_{\text{total}}$ to be one. 
	The single queue mechanisms all outperform the straightforward two queue mechanism. Note that $\Var[T]^{\text{2Q}}$ reaches minimum at
	\begin{equation}
	 \frac{\lambda_1}{\lambda_{\text{total}}} = 2\sqrt{33} - 11 \approx 0.489.
	\end{equation}
	
	It is also worth noting that though it is straightforward to argue FIFO and packing outperform LIFO service order, their variances do \emph{not} differ much. Thus, \textbf{the optimal matchmaking mechanism is central queue in FIFO/packing service order}, but the whole family of single queue matchmaking mechanisms are \emph{insensitive} to service orders (and hence small network latency!) in the sense of reasonable waiting time variances.
	
	\vspace{0.2cm}
	\noindent \textbf{Remark}: For games with more players, say $k$v$k$ battle games, it is more complicated to design the optimal matchmaking mechanism. However, the intuition is that packing would be a good choice when $k$ is small. A relatively large $k$ may display the problem's combinatorial nature in that we need to solve a subset sum problem to decide which $k$ players to pack together whenever a new player arrives.
	\vspace{0.2cm}
	
	Having derived the optimal matchmaking mechanism, we are ready to examine different rule designs' impacts on the matchmaking mechanism through the expected waiting time. We are especially interested in two kinds of rule designs for 2v2 battle game: allowing players to choose sides, and allowing players to join 2 queues simultaneously in the 2-queue matchmaking mechanism.

	\section{Static Rule Design: Choosing Sides}
	\label{sec:static}
	In some 2v2 games, there could be possible differences in all aspects of figures between the two battle sides. Players may prefer one side than another. This leads to the first rule design of our interest: it would be nice to allow players to choose sides as this will obviously improve the players' satisfaction to the game. However, this rule may inevitably increase the expected waiting time for the game (Lemma \ref{thm_2v2} \emph{won't} hold with this additional rule!). We investigate these trade-offs in this section. 
	
	In the basic model analyzed in the previous section, there only exist \textit{finite} possible states. However, once the side selection is allowed, the state space can go to \textit{infinity}. Choosing sides sophisticates the rule design. 
	This warrants us exploring a simple but representative approximation of the 2v2 game. To obtain useful insights for rule design through a neat analysis, in this work, we select the 1v1 game to approximate the 2v2 game. The first subsection demonstrates the conditions under which such approximation is accurate. Then, we present insights obtained from the approximated game, which is crucial for rule designs.

	\subsection{Approximation for 2v2 Game with Packing}
	
	
	The 2v2 game with packing service order can be viewed as an 1v1 game by counting each package of 2 players as a single arrival. However, such 1v1 game approximation will \emph{not} lead to a CTMC. This is because packing two individual players makes the inter-arrival time distribution non-exponential.
	
	By normalizing $\lambda_\text{total}$ to be 1, we plot the histograms for the inter-arrival time of 1v1 game approximation with diverse $\lambda_1$'s in Fig. \ref{fig:approx}. We also compare the histograms with Poisson arrival of the same total arrival rate. We can observe that, the only hurdle preventing the distribution to be exponential is the arrival of individual players. As $\lambda_1$ increases, the 1v1 game approximation becomes more inaccurate.
	
	Nonetheless, in the subsequent analysis, we stick to the 1v1 battle game approximation for the 2v2 battle game to obtain more insights through analytical studies. We want to emphasize that though our conclusions are only valid for the case when the game attracts significantly more team arrivals than individual arrivals, they provide useful intuition for other cases as well.
	
	
	\subsection{Impacts of Proportions Having Preferred Side}
	\label{sec:hurt}
	The analysis on 1v1 game approximation reveals that the proportion of players preferring one side, referred by the players' structure in the following sections,  determines the consequence of allowing side selection. 
	
\begin{figure}[t]
		\centering
		\begin{tikzpicture}
		\graph[->/.tip=stealth,
		grow right=1.3cm,
		nodes={circle,fill=black!15,inner sep=0,minimum size=8mm}]
		{
			{
				[edges={bend left,"$\lambda_A$" above}]
				c[as=$\cdots$,fill=black!5] -> b[as=-2] -> a[as=-1] -> 0 -> 1 -> 2 -> 3[as=$\cdots$,fill=black!5];
			};
			{
				[edges={bend left,"$\lambda_B$" below}]
				3 -> 2 -> 1 -> 0 -> a -> b -> c;
			};
		};
		\end{tikzpicture}
		\caption{The CTMC for allowing all players to choose sides.}
		\label{2sidequeue}
	\end{figure}
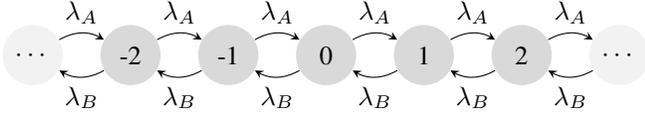

	We first demonstrate the effects of the players' structure by the extreme case where we allow and require all the players to choose a side of their favorites: namely either side A or side B. Denote the arrival rate of players choosing side A by $\lambda_A$, and that of players choosing side B by $\lambda_B$. To figure out the CTMC, it suffices to identify its states: using integer $i$ is enough. For positive $i$, state $i$ implies there are $i$ players choosing side A in the queue. For negative $i$, state $i$ implies $-i$ players choosing side B. Figure \ref{2sidequeue} plots such a CTMC.
	
	Unfortunately, this CTMC is quite classical in that it is transient when $\lambda_A \neq \lambda_B$, and null recurrent when $\lambda_A = \lambda_B$. Hence, even in the latter case, one can show that the expected waiting time for all players is \emph{unbounded}. Hence, \emph{game designers cannot afford to allow all players to choose sides freely}. 
	
\begin{figure}[t]
		\centering
		\begin{tikzpicture}
		\tikzstyle{every node}=[font=\footnotesize]
		\node[circle,fill=black!15,inner sep=0,minimum size=8mm] (4) at (3.9,-1.95) {$1'$};
		\graph[->/.tip=stealth,
		grow right=1.3cm,
		nodes={circle,fill=black!15,inner sep=0,minimum size=8mm}]
		{
			[edges=bend left]
			{
				c[as=$\cdots$,fill=black!5] ->["$\lambda_A + \lambda_C$" above] b[as=-2] ->["$\lambda_A + \lambda_C$" above] a[as=-1] ->["$\lambda_A + \lambda_C$" above] 0 ->["$\lambda_A$" above] 1 ->["$\lambda_1$" above] 2 ->["$\lambda_A$" above] 3[as=$\cdots$,fill=black!5];
			};
			{
				[edges="$\lambda_B + \lambda_C$" below]
				3 -> 2 -> 1 -> 0;
			};
			{
				[edges="$\lambda_B$" below]
				0 -> a -> b -> c;
			};
			0 ->["$\lambda_C$" right, bend left=5, near end] (4) ->["$\lambda_A + \lambda_B + \lambda_C$" left, bend left=5, near start] 0;
		};
		\end{tikzpicture}
		\caption{The CTMC with choice-free players.}
		\label{2sidequeue2}
	\end{figure}
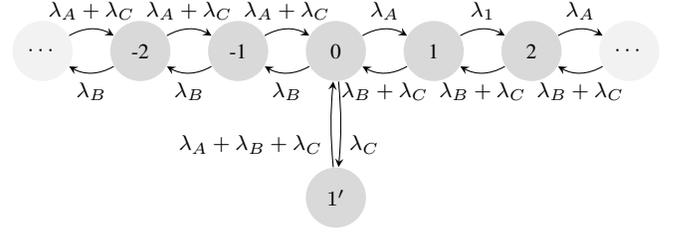

	However, introducing a small portion of choice-free players (who are indifferent in choosing sides) into the game can inverse the conclusion. Denote the arrival rate of choice-free players by $\lambda_C$. Figure. \ref{2sidequeue2} plots the new CTMC.
	\begin{fact}
	The CTMC in Fig. \ref{2sidequeue2} is positive recurrent iff
	\begin{equation}\label{lambda_conditon}
	    \lambda_C>|\lambda_A-\lambda_B|.
	\end{equation}
	\end{fact}
	
	\noindent\textbf{Remark}: This is a straightforward yet remarkable fact. If by design the two battle sides are balanced, i.e., $\lambda_A$ and $\lambda_B$ do not differ much, then the game can afford allowing more players to choose sides. In contrast, if the game is not carefully designed, we may have to force many players to take the side they don't like. This will further reduce their satisfaction to the game: a disaster to the game designer.
	
	The above remark indicates that \textit{the valuable players include not only the choice-free players but also the players balancing the proportions of the two sides}. Denote the waiting time for players choosing side A, players choosing side B, and choice-free players by $T_A$, $T_B$, and $T_C$, respectively. Solving the CTMC in Fig. \ref{2sidequeue2} yields (see Appendix \ref{app:static} for more details)
	\begin{align}
		\E[T_A] &= \frac{(\lambda_B + \lambda_C)\pi_0}{(\lambda_B + \lambda_C - \lambda_A)^2},\label{eqTa}\\
		\E[T_B]& = \frac{(\lambda_A + \lambda_C)\pi_0}{(\lambda_A + \lambda_C - \lambda_B)^2},\label{eqTb}\\
		\E[T_C]& = \frac{\pi_0}{\lambda_A + \lambda_B + \lambda_C},\label{eqTc}
	\end{align}
	where $\pi_0$ denotes the stationary probability at state $0$. Denote $\lambda_{\text{total}}$ as the total arrival rate $\lambda_A + \lambda_B + \lambda_C$ into the system, we have
	\begin{equation}
		\pi_0 = \left(\frac{\lambda_A}{\lambda_B \!+\! \lambda_C\! -\! \lambda_A} \!+\! \frac{\lambda_B}{\lambda_A \!+\! \lambda_C\! -\! \lambda_B} + \lambda_C\lambda_{\text{total}}^{-1} + 1\right)^{-1}\!\!\!.\nonumber
	\end{equation}

	With Eq. (\ref{eqTa})-(\ref{eqTc}), we can further obtain the expected waiting time for all players:
	\begin{equation}
		\begin{aligned}
			\E[T] &= \frac{1}{2(\lambda_A+\lambda_C-\lambda_B)}+\frac{1}{2(\lambda_B+\lambda_C-\lambda_A)}\\
			&\quad +\frac{1}{\lambda_A+\lambda_B+\lambda_C} -\frac{1}{2\lambda_C}\\
			&\quad -\frac{\lambda_A+\lambda_B+2 \lambda_C}{2(2\lambda_A \lambda_B+\lambda_A\lambda_C+\lambda_B\lambda_C+\lambda_C^2)}.
		\end{aligned}
		\label{etside}
	\end{equation}

\begin{figure}[t]
	    \centering
	    \includegraphics[width=2.6in]{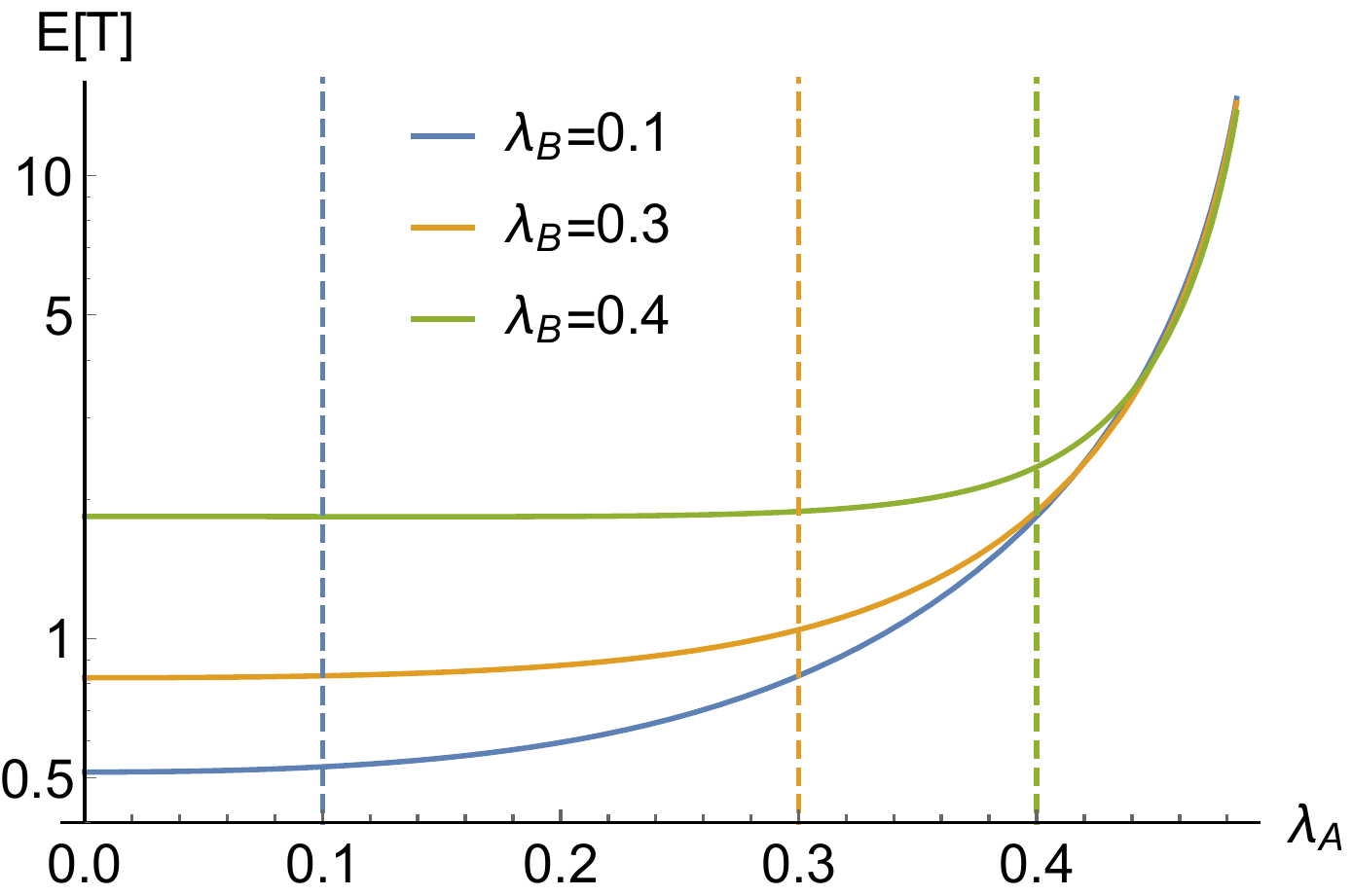}
	    \caption{The expected waiting time with normalized $\lambda_{\text{total}}$.}
	    \label{sidegraph}
	\end{figure}

	To decipher Eq.\eqref{etside}, we normalize $\lambda_{\text{total}}$ to be 1. 
	Figure \ref{sidegraph} examines how $\E[T]$ varies with $\lambda_A$ and $\lambda_B$. With normalized $\lambda_{\text{total}}$, for any fixed $\lambda_B$, $\E[T]$ hardly changes with  $\lambda_A$ as long as $\lambda_A < \lambda_B$. In contrast, when $\lambda_A > \lambda_B$, $\E[T]$ increases sharply. Due to the symmetric nature of $\lambda_A$ and $\lambda_B$ to $\E[T]$, we conclude that $\max(\lambda_A, \lambda_B)$ dominates $\E[T]$. That is, the benefit of being choice-free depends on the \emph{imbalance} between $\lambda_A$ and $\lambda_B$. This can be further exemplified by the improvement factor: we define improvement factor $q$ as follows:
	\begin{equation}
	    q = \frac{\E[T_C]}{\E[T_B]} = \frac{(\lambda_{\text{total}} - 2\lambda_B)^2}{\lambda_{\text{total}}(\lambda_{\text{total}} - \lambda_B)}.
	\end{equation}
	Given $\lambda_{\text{total}}=1$, we can further study the first order derivative of $q$ with respect to $\lambda_B$:
    \begin{equation}
        \frac{\partial q}{\partial \lambda_B} = \frac{1}{(1-\lambda_B)^2} - 4 < 0.
    \end{equation}
    This derivative is always negative in that we assume $\lambda_B < 0.5 \lambda_\text{total}$. Hence, as $\lambda_B$ increases (the two sides become more symmetric), the improvement factor decreases, and decreases \emph{slower}!
    
\begin{figure}[t]
	    \centering
	    \includegraphics[width=2.6in]{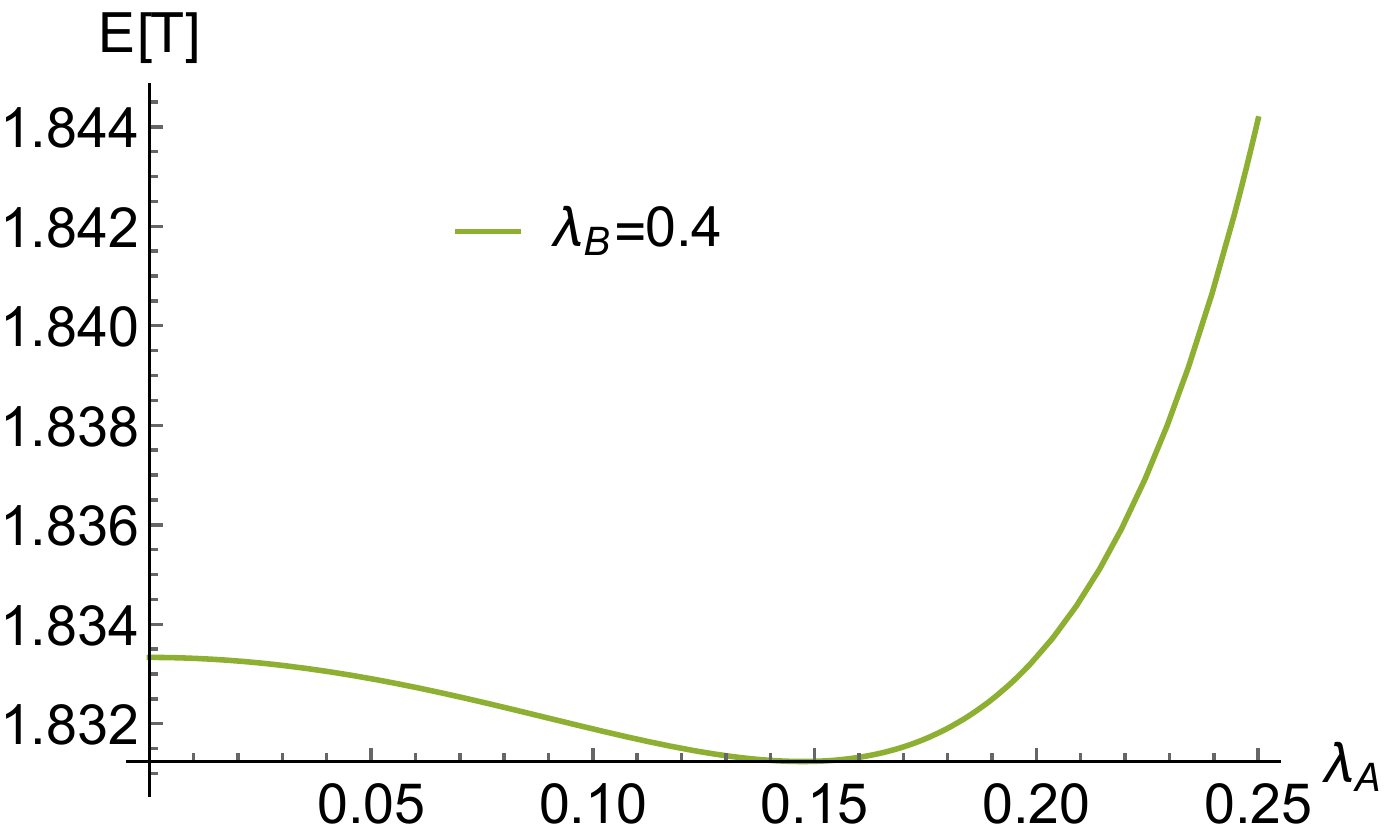}
	    \caption{$\E[T]$ with normalized $\lambda_{\text{total}}$ and $\lambda_B = 0.4$.}
	    \label{sidegraph_zoom_in}
	\end{figure}

    
	However, $\E[T]$ does not display such monotonicity with respect to $\lambda_B$. Zooming in Fig. \ref{sidegraph}, we plot Fig. \ref{sidegraph_zoom_in} to highlight that though $\E[T]$  is generally insensitive to the changes in $\lambda_A$ when $\lambda_A<\lambda_B$, increasing $\lambda_A$ could first help reduce the mean waiting time, then hurt the system performance! 

	\subsection{Guideline for Rule Design}
	To sum up, it is crucial to evaluate players' preference distribution before the rule design. With a more symmetric player preference distribution, the game could afford more players to choose sides according to their preferences. However, this issue raises a very important future work: players' preference may change over time, and thus the ideal rule game could be somewhat dynamic. Hence, it is important to establish an adaptive rule design to constantly track the changes in players' preference distribution. Also, the mechanism design of attracting players balancing the sides also deserves carefully explorations. 
	
	\section{Dynamic Rule Design: Climbing the Ladder}
	\label{sec:dynamic}
	We have focused on the static analysis for the 2v2 battle game in the sense that we assume all players are identical across time. In this section, we consider a dynamic setting to allow the players to have different levels. This is true for many games: players start from the beginner level and can only play in a junior zone. When their skills climbing up to a higher level, they are allowed to play in the advanced zone\footnote{One concrete example could be the game \textit{Majsoul} \cite{ref_majsoul}. There are 5 arenas: Bronze, Silver, Gold, Jade and Throne Arena. Players are allowed to battle in different arenas according to their levels in the game.}. Though in practice, there could be more zones for a game. We restrict ourselves to the two-zone scenario for analytical solutions.
	
	\subsection{Battle in Both Zones: the Formulation}
	One straightforward rule design problem could be to decide the level to distinguish the players for two zones. However, such a decision making problem heavily depends on concrete game setting. In this section, we turn to another decision making problem: some games allow the experienced players to battle in the junior zone while others don't. In terms of the expected waiting time, how much would it help to allow the experienced players to \emph{battle in both zones}?
	
	We follow the 1v1 game approximation in this section to enable us focus on understanding the dynamic rule design's impact on matchmaking. In such a setting, there are 3 types of arrivals, green players for Zone A (the junior zone) arrives with rate $\lambda_A$, experienced players only for Zone B (the advanced zone) arrives with rate $\lambda_B$, and experienced players for both zones (choice-free players) arrives with rate $\lambda_C$.
	
	
	
\begin{figure}[t]
		\centering
		\begin{tikzpicture}[->/.tip=stealth,nodes={inner sep=0,minimum size=8mm}]
		\node[circle,fill=black!15] (0) at (0, 0) {0};
		\node[circle,fill=black!15] (A) at (2.3, 1.5) {A};
		\node[circle,fill=black!15] (B) at (2.3, -1.5) {B};
		\node[circle,fill=black!15] (C) at (-2.3, 0) {C};
		\node[circle,fill=black!15] (AB) at (4.6, 1.5) {AB};
		\node[circle,fill=black!15] (BA) at (4.6, -1.5) {BA};
		\draw (0) edge[->,"$\lambda_A$" above left, bend left] (A);
		\draw (B) edge[->,"$\lambda_A$" above] (BA);
		\draw (BA) edge[->,"$\lambda_A$" below, bend left] (B);
		\draw (0) edge[->,"$\lambda_B$" above right] (B);
		\draw (A) edge[->,"$\lambda_B$" above, bend left] (AB);
		\draw (AB) edge[->,"$\lambda_B$" below] (A);
		\draw (0) edge[->,"$\lambda_C$" below, bend left] (C);
		\draw (A) edge[->,"$\lambda_A + \lambda_C$" below right] (0);
		\draw (AB) edge[->,"$\lambda_A + \lambda_C$" below right, near start] (B);
		\draw (B) edge[->,"$\lambda_B + \lambda_C$" below left, bend left] (0);
		\draw (BA) edge[->,"$\lambda_B + \lambda_C$" above right, near start] (A);
		\draw (C) edge[->,"$\lambda_A + \lambda_B + \lambda_C$" above, bend left] (0);
		\end{tikzpicture}
		\caption{CTMC for dynamic rule design in FIFO order.}
		\label{1v1sfifo}
	\end{figure}
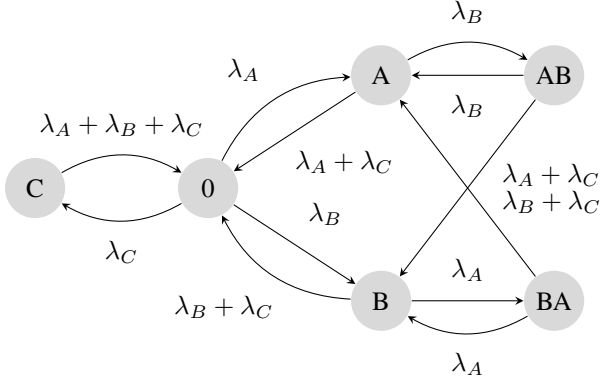

	However, this is not enough to formulate the CTMC. It remains undecided when a choice-free experienced player joins the two queues and finds that both queues are non-empty. For simplicity, we employ the FIFO service order to deal with such situations. That is, match the new arrival with the player who arrives first (In fact, before the new arrival, there are exactly two players in the system, and one in each queue). We use \emph{state 0} to indicate there is no player in either queue, and use \emph{states A, B, and C} to indicate from \emph{state 0}, the new arrival is a green player for Zone A, experienced player for Zone B, and choice-free player for both zones, respectively. We denote the state when the player queueing in Zone A arrives first by \emph{state AB}. In contrast, we denote the corresponding state when the player queueing in Zone B arrives first by \emph{state BA}. With these notations, we plot the CTMC in Fig. \ref{1v1sfifo}.
	
	Solving this CTMC, Little's Law dictates that (see Appendix \ref{app:dynamic} for more details)
	\begin{equation}
		\E[T_A]\! =\! \frac{2\lambda_A \lambda_B \!+\! \lambda_A \lambda_C \!+\! 2\lambda_B^2 \!+\! 4\lambda_B \lambda_C \!+\! \lambda_C^2}{(\lambda_A\!+\!\lambda_C) (\lambda_B\!+\!\lambda_C) (\lambda_A\!+\!\lambda_B\!+\!\lambda_C)}\pi_0,
		\label{1v1set1fifo}
	\end{equation}
	\begin{equation}
		\E[T_B] \!=\! \frac{2\lambda_A \lambda_B \!+\! \lambda_B \lambda_C \!+\! 2\lambda_A^2 \!+\! 4\lambda_A \lambda_C \!+\! \lambda_C^2}{(\lambda_A\!+\!\lambda_C) (\lambda_B\!+\!\lambda_C) (\lambda_A\!+\!\lambda_B\!+\!\lambda_C)}\pi_0,
	\end{equation}
	\begin{equation}
		\E[T_C] = \frac{\pi_0}{\lambda_A + \lambda_B + \lambda_C},
	\end{equation}
	where $\pi_0$ is the stationary distribution for being at \emph{state 0}. Using the law of total probability, we can obtain the expected waiting time for all arrivals:
	\begin{equation}
	   \E[T] 
			= \frac{\lambda_A \E[T_A] + \lambda_B \E[T_B] + \lambda_C \E[T_C]}{\lambda_A + \lambda_B + \lambda_C}.
	\end{equation}

	
	\subsection{Benefits of Being Choice-Free}
	
	To understand the benefits for experienced players of being choice-free, we define the improvement factor $q$ as follows:
	\begin{equation}
		q \!=\! \frac{\E[T_C]}{\E[T_B]} \!=\! \frac{(\lambda_A+\lambda_C) (\lambda_B+\lambda_C)}{2\lambda_A^2 \!+\! 2\lambda_A \lambda_B \!+\! 4\lambda_A \lambda_C \!+\! \lambda_B \lambda_C \!+\! \lambda_C^2}.
	\end{equation}
	Obviously, $q$ measures how much it saves the experienced players to start a game by joining two queues, instead of sticking to the advanced zone. To obtain more insights, note that when $\lambda_A = \lambda_B$, we have
	\begin{equation}
		q = \frac{\lambda_B + \lambda_C}{4\lambda_B + \lambda_C}.
	\end{equation}
	That is, when $\lambda_A = \lambda_B$, the benefit for the first experienced player to switch to a choice-free player is to save 75\% expected waiting time! As the number of choice-free players increases, such benefit diminishes.

\begin{figure}[t]
	    \centering
	    \includegraphics[width=3in]{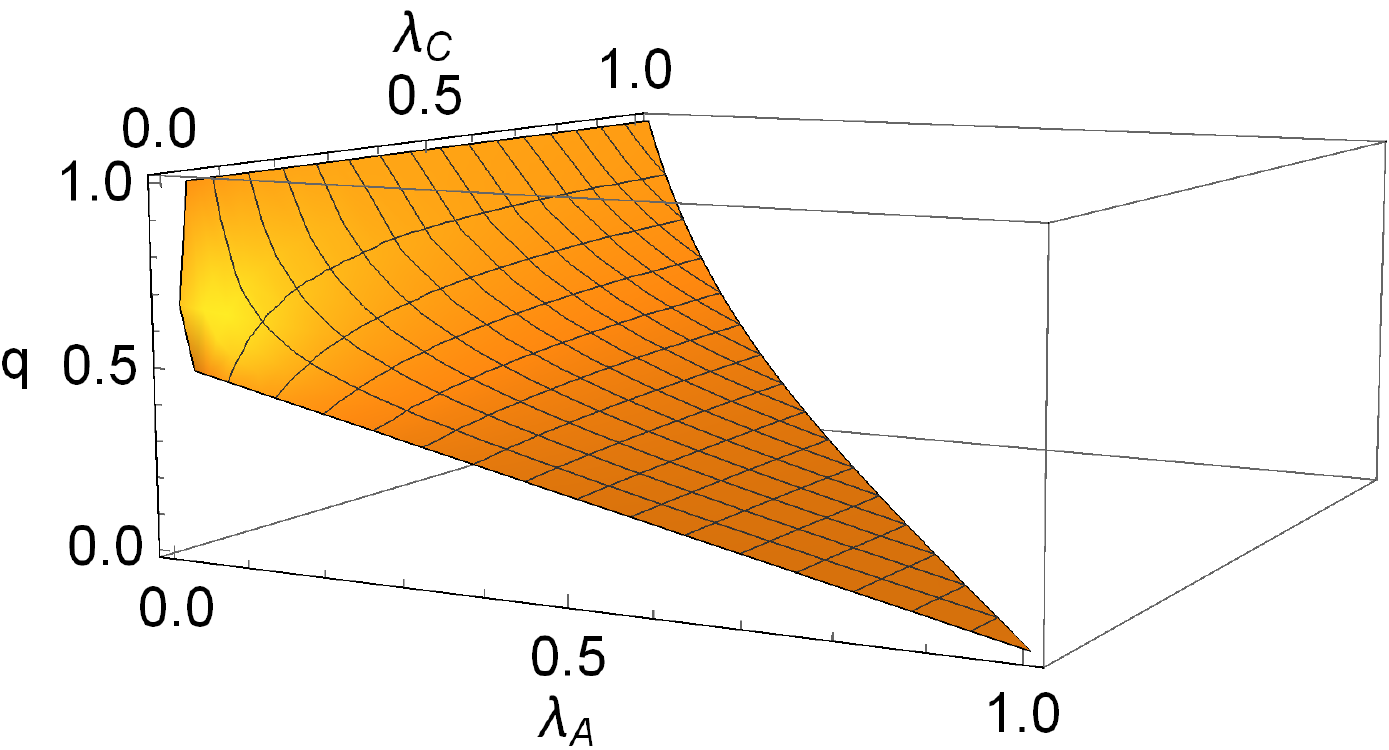}
	    \caption{The improvement factor $q$ vs. $\lambda_A$ and $\lambda_C$.}
	    \label{fig:2_zones}
	\end{figure}

	As Fig. \ref{fig:2_zones} illustrates, when $\lambda_C=0$, and $\lambda_\text{total}=1$,
	\begin{equation}
	    q=0.5(1-\lambda_A),
	\end{equation}
	which implies that the initial benefit of being choice-free solely depends on $\lambda_A$! When the junior zone is already crowded compared to the advanced zone, then there is no incentive for being choice-free. 
	
	\begin{figure}[t]
	    \centering
	    \includegraphics[width=3in]{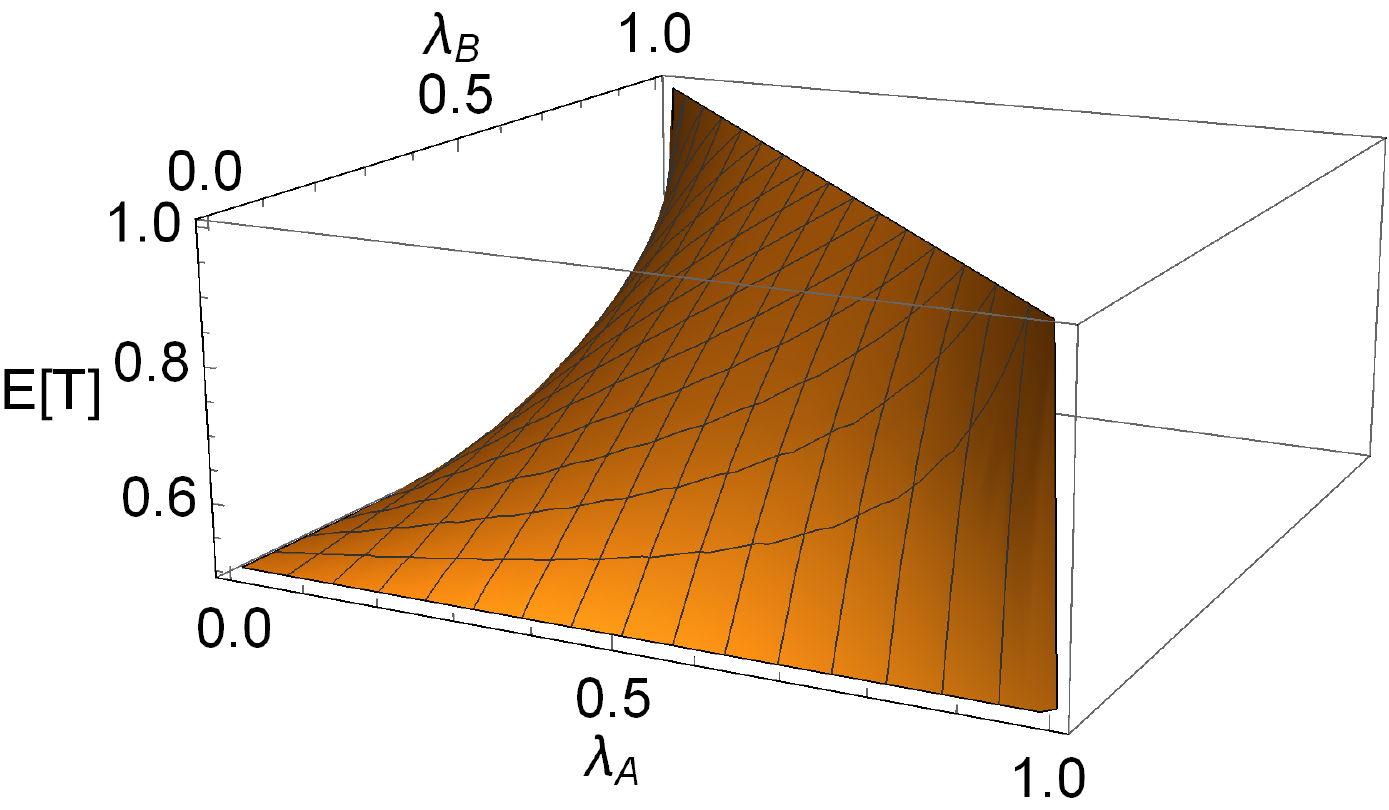}
	    \caption{$\E[T]$ vs. $\lambda_A$ and $\lambda_B$ with normalized $\lambda_{\text{total}}$.}
	    \label{fig:2_zones_et}
	\end{figure}

	Besides the intuitions, we can further conclude that $q$ increases rapidly (and hence the benefit decreases rapidly) as $\lambda_A$ decreases (fewer junior players). In fact, the maximal benefit of being choice-free could save the players up to 90\% of expected waiting time when $\lambda_A=0.8$ and $\lambda_B=0.2$.
	
	The system also benefits from the existence of choice-free players, as shown in Fig.~\ref{fig:2_zones_et}. The more the choice-free players more, the lower the expected total waiting time is. However, the marginal system value of additional choice-free player decreases. Nevertheless, the proportion of choice-free players in the total population determines whether the game designers should allow the players to select the battle zone according to their own preferences.

	
	
	\section{Conclusion}
	\label{sec:conclusion}
	In this paper, we focus on the optimal matchmaking mechanisms for 2v2 battle game. Using the 1v1 game approximation to the packing service order, we analyze both the static rule design as well as the dynamic rule design's impact on the matchmaking mechanisms. We submit that in both cases, choice-free players are vital to the system: being choice-free benefits themselves as well as the whole system.
	
	
	Much remains unknown. The CTMC for games with more players are more complicated, and it is hard to obtain analytical insights. Also, in this paper, our goal for the game rule design is to minimize the expected waiting time. However, in practice, many more aspects concerning the player's satisfaction to the game need to be carefully examined.
	
	It would be also interesting to consider more dynamic scenario when players' preference distribution, and players' level distribution change over time. This inspires us to construct more adaptive rule designs in the future.

	\section{Appendix}
	
	\subsection{$\E[T]$ for static rule design}
	\label{app:static}
	Observe the time-reversibility equations:
	\begin{equation*}
	    \left\{
    	\begin{aligned}
    		\lambda_A\pi_0 &= (\lambda_B + \lambda_C) \pi_1  \\
    		\lambda_A\pi_1 &= (\lambda_B + \lambda_C) \pi_2  \\
    		& \hspace{-0.08333em}\cdots  \\
    		\lambda_B\pi_0 &= (\lambda_A + \lambda_C) \pi_{-1}  \\
    		\lambda_B\pi_{-1} &= (\lambda_A + \lambda_C) \pi_{-2}  \\
    		& \hspace{-0.08333em}\cdots  \\
    		\lambda_C \pi_0 &= (\lambda_A + \lambda_B + \lambda_C) \pi_{1'}. 
    	\end{aligned}
	    \right.
	\end{equation*}
	We define the following two series for $z$-transforms:
	\begin{align}
		\widehat{N_1}(z) &= \sum\nolimits_{i=1}^{\infty} \pi_i z^i, \\
		\widehat{N_2}(z) &= \sum\nolimits_{i=1}^{\infty} \pi_{-i} z^i.
	\end{align}
	The time-reversibility equations imply
	\begin{align}
		\lambda_A z \widehat{N_1}(z) + \lambda_A \pi_0 z &= (\lambda_B + \lambda_C) \widehat{N_1}(z), \\
		\lambda_B z \widehat{N_2}(z) + \lambda_B \pi_0 z &= (\lambda_A + \lambda_C) \widehat{N_2}(z).
	\end{align}
	Having obtained the closed form expressions for $\widehat{N_1}(z)$ and $\widehat{N_2}(z)$, we can derive the $z$-transform for the stationary distributions $\widehat{N}(z)$ via
	\begin{equation}
			\widehat{N}(z) = \widehat{N_1}(z) + \widehat{N_2}(z) + \pi_{1'}z + \pi_0 .
	\end{equation}
	We submit that
	\begin{align}
			\E[T_A] &= \left(\widehat{N_1}'(1) + \widehat{N_1}(1) + \pi_0\right) \cdot \frac{1}{\lambda_B + \lambda_C}, \\
		\E[T_B] &= \left(\widehat{N_2}'(1) + \widehat{N_2}(1) + \pi_0\right) \cdot \frac{1}{\lambda_A + \lambda_C}, \\
			\E[T_C]& = \pi_0(\lambda_A + \lambda_B + \lambda_C)^{-1}.
	\end{align}
	Law of total probability leads to $\E[T]$:
	\begin{equation}
		\E[T]= \frac{\lambda_A \E[T_A] + \lambda_B \E[T_B] + \lambda_C \E[T_C]}{\lambda_A + \lambda_B + \lambda_C}.
	\end{equation}
	Standard mathematical manipulation yields the desirable results. \hfill$\blacksquare$
	
	\subsection{$\E[T]$ for dynamic rule design}
	\label{app:dynamic}
    Denoting $\lambda_\text{total}=\lambda_A+\lambda_B+\lambda_C$, it suffices to identify the balance equations for the CTMC are as follows:
    \begin{equation*}
        \left\{
    	\begin{aligned}
    		\lambda_\text{total}\cdot \pi_0 &= \lambda_\text{total}\cdot\pi_C \!+\! (\lambda_A \!+\! \lambda_C) \pi_A \!+\! (\lambda_B \!+\! \lambda_C) \pi_B  \\
    		\lambda_\text{total}\cdot\pi_A &= \lambda_A \pi_0 + \lambda_B \pi_{AB} + (\lambda_B + \lambda_C) \pi_{BA}  \\
    		\lambda_\text{total}\cdot \pi_B &= \lambda_B \pi_0 + \lambda_A \pi_{BA} + (\lambda_A + \lambda_C) \pi_{AB}  \\
    		\lambda_\text{total}\cdot \pi_{AB} &= \lambda_B \pi_A  \\
    		\lambda_\text{total}\cdot \pi_{BA} &= \lambda_A \pi_B  \\
    		\lambda_\text{total}\cdot \pi_C &= \lambda_C \pi_0. 
    	\end{aligned}
    	\right.
    \end{equation*}
	It is straightforward to verify the following stationary distributions are the solution to the above balance equations:
	\begin{align}
		\pi_A \ \ &= \lambda_A \pi_0(\lambda_A+\lambda_C)^{-1}, \\
		\pi_B \  \ &= \lambda_B \pi_0(\lambda_B+\lambda_C)^{-1}, \\
		\pi_{AB} &= \lambda_A \lambda_B \pi_0(\lambda_A+\lambda_C)^{-1} \lambda_\text{total}^{-1}, \\
		\pi_{BA} &= \lambda_A \lambda_B \pi_0(\lambda_B+\lambda_C)^{-1} \lambda_\text{total}^{-1}, \\
		\pi_C \ \ &= \lambda_C \pi_0 \lambda_{\text{total}}^{-1}.
	\end{align}
	We take $\E[T_A]$ as an example to illustrate how to obtain our desirable results. Let $f_i$ denote the expected remaining waiting time for junior players (with arrival rate $\lambda_A$) when the system is at state $i$. With such definitions, we can express $\E[T_A]$ as follows:
	\begin{equation}
		\E[T_A] = \pi_0 f_A + \pi_2 f_{BA}.
	\end{equation}
	Hence, it suffices to derive $f_A$ and $f_{BA}$. By definition of the $f_i$'s, law of total probability yields
	\begin{align}
		f_A \ \  &= \lambda_\text{total}^{-1} + \lambda_B f_{AB}\lambda_\text{total}^{-1}, \\
		f_{AB} &= \lambda_\text{total}^{-1} + \lambda_B f_{A}\lambda_\text{total}^{-1}, \\
		f_{BA} &= \lambda_\text{total}^{-1} + (\lambda_B + \lambda_C) f_A\lambda_\text{total}^{-1}.
	\end{align}
	Solving this system of equations immediately leads to $\E[T_A]$. Other results can be derived in the same routine. \hfill$\blacksquare$



\begin{thebibliography}{99}
		
		\bibitem{ref_lol}
		 ``League of Legends'', https://euw.leagueoflegends.com/.
		 
		\bibitem{ref_trace}
		 Maxime Véron, Olivier Marin, and Sébastien Monnet, ``Matchmaking in multi-player on-line games: studying user traces to improve the user experience'', {\it Proceedings of Network and Operating System Support on Digital Audio and Video Workshop}, ACM, 2014.
		
		\bibitem{ref_latency_sensitive}
		 Sharad Agarwal, and Jacob R. Lorch, ``Matchmaking for online games and other latency-sensitive P2P systems'', {\it ACM SIGCOMM Computer Communication Review}, Vol. 39, No. 4, ACM, 2009.
		
		\bibitem{ref_switchboard}
		 Justin Manweiler, et al, ``Switchboard: a matchmaking system for multiplayer mobile games'', {\it Proceedings of the 9th international conference on Mobile systems, applications, and services}, ACM, 2011.
		
		\bibitem{ref_haste}
		 Yuval Emek, Shay Kutten, and Roger Wattenhofer, ``Online matching: haste makes waste!'', {\it Proceedings of the forty-eighth annual ACM symposium on Theory of Computing}, ACM, 2016.
		 
		\bibitem{ref_game_structure_sensitive}
		 Mateusz Myślak, and Dominik Deja, ``Developing game-structure sensitive matchmaking system for massive-multiplayer online games'', {\it International Conference on Social Informatics}, Springer, Cham, 2014.
		 
		\bibitem{ref_foundation}
		 Josh Alman, and Dylan McKay, ``Theoretical foundations of team matchmaking'', {\it Proceedings of the 16th Conference on Autonomous Agents and MultiAgent Systems}, International Foundation for Autonomous Agents and Multiagent Systems, 2017.
		 
		
		\bibitem{ref_little}
		 J.D.C. Little, ``A proof of the queueing formula $L=\lambda W$'', {\it Operations Research}, 9:383-387, 1961.
		 
		\bibitem{ref_cr}
		 ``Clash Royale'', https://clashroyale.com/.
		 
		\bibitem{ref_majsoul}
		 ``Majsoul'', https://www.majsoul.com/. 
		
	\end{thebibliography}
\end{document}